\documentclass{JFM-FLM_Au}
\hypersetup{citecolor=blue,linkcolor=blue}
\usepackage[T1]{fontenc}
\usepackage{amsmath,amssymb}
\usepackage[mathlines]{lineno}

\makeatletter
\renewcommand{\ps@headings}{%
  \let\@mkboth\markboth
  \def\@oddhead{\hfill{\itshape\@righttitle}\hfill}%
  \def\@evenhead{\hfill\itshape\@lefttitle\hfill}%
  \def\@oddfoot{\hfil\thepage\hfil}%
  \def\@evenfoot{\hfil\thepage\hfil}%
  \def\sectionmark##1{\markboth{##1}{}}%
  \def\subsectionmark##1{\markright{##1}}%
}
\makeatother
\newcommand{\ii}{\mathrm{i}}
\newcommand{\sgn}{\operatorname{sgn}}

\lefttitle{P. Simson}
\righttitle{A real-variable unidirectional reduction of deep-water gravity waves}

\title{A real-variable unidirectional reduction of deep-water gravity waves
}

\author{P\"aivo Simson\aff{1}}

\affiliation{\aff{1}Laboratory of Solid Mechanics, Department of Cybernetics, Tallinn University of Technology, Tallinn, Estonia
}

\corresau{P\"aivo Simson, \email{paivo.simson@taltech.ee}}

\begin{document}

\maketitle

\begin{abstract}
A unidirectional reduction of the deep-water surface gravity wave problem is derived in physical space using real variables. By employing a near-identity canonical transformation, cubic interactions are eliminated from the Hamiltonian, resulting in a bidirectional system analogous to the 1D Zakharov equation. An exact factorization of the linear dynamics then selects the rightward-propagating branch, which is carried to second order in the wave steepness, yielding a single nonlocal evolution equation with full linear dispersion. The model admits an exact monochromatic solution whose reconstructed surface elevation is the third-order Stokes wave, and a multiple-scales analysis recovers the Dysthe envelope equation complete with its nonlocal mean-flow term. The equation also admits exact breather solutions, whose profiles agree closely with those of the super compact equation. Dropping the sub-leading nonlinear terms that vanish on the resonant manifold provides a more compact variant suitable for analytical study. Numerical comparisons show that both formulations track the fully nonlinear Euler dynamics closely through modulational-instability recurrence and broadband focusing up to moderate wave steepness.

\end{abstract}

\begin{keywords}

\end{keywords}

\section{Introduction}\label{sec:intro}

The modern mathematical treatment of deep-water surface gravity waves stems largely from the foundational step taken by \citet{Zakharov1968}, who demonstrated that the surface elevation $\eta$ and the surface velocity potential $\phi$ form canonically conjugate variables, with the dynamics derivable from a single Hamiltonian functional -- the total energy of the fluid. Subsequently, in the same study, Zakharov introduced complex canonical variables in Fourier space and derived the nonlinear Schr{\"o}dinger equation (NLS) as the leading-order envelope dynamics of a slowly modulated wave packet.

The Hamiltonian framework supported decades of subsequent work in Fourier-space canonical variables. A striking result by \citet{DyachenkoZakharov1994} revealed that the four-wave interaction coefficient vanishes identically on the nontrivial resonant manifold for one-dimensional deep-water gravity waves. This unexpected cancellation, combined with Krasitskii's systematic normal-form analysis \citep{Krasitskii1994}, paved the way for the compact \citep{DyachenkoZakharov2011,DyachenkoZakharov2012} and super compact \citep{DyachenkoKachulinZakharov2017} equations: unidirectional reductions of the governing equations in complex normal variables.
In parallel, \citet{DyachenkoEtAl1996} developed conformal-mapping methods that solve the water-wave problem numerically without truncation in $\epsilon$.

The modulation theory of weakly nonlinear wavepackets was developed somewhat independently of the Hamiltonian framework, through direct asymptotic expansions on the boundary-value problem. \citet{Dysthe1979} extended the nonlinear Schr{\"o}dinger equation to higher order in steepness; later refinements \citep{TrulsenDysthe1996, CraigGuyenneSulem2010, GramstadTrulsen2011} 
extended the framework to broader bandwidth and Hamiltonian form. These envelope equations remain the standard for practical analysis, though they are inherently localized around a single carrier wavenumber.

A distinct line of work, pioneered by \citet{Matsuno1992, Matsuno1993}, formulates the problem directly in physical variables using the Hilbert transform. This approach yields broadband, bidirectional nonlocal equations that serve as the deep-water analogues of the shallow-water Boussinesq equations. Independently, \citet{CraigSulem1993} developed a more general framework based on the Taylor expansion of the Dirichlet-to-Neumann operator. The real-variable Hilbert-transform formulation has been revisited recently by \citet{Krechetnikov2024}, who derives a bidirectional $\mathcal{O}(\epsilon^2)$ equation in the surface velocity potential analogous in spirit to that of \citet{Matsuno1993} and uses it to streamline the derivation of NLS. To our knowledge, a unidirectional reduction in real-variable operator form carried through to Dysthe order has not previously appeared in the literature. This gap was originally pointed out by \citet{Matsuno1992} and reiterated by \citet{Choi_1995}.

The present paper develops such a reduction. We start from the Hamiltonian formulation and perform a canonical transformation that eliminates the cubic Hamiltonian, yielding equations of motion in real variables $(u,v)$. An exact factorization of the linear bidirectional equation then defines the pseudo-differential operator $A$, with which the dynamics is reduced to the rightward-propagating branch through $\mathcal{O}(\epsilon^2)$. Discarding the terms that vanish on the resonant manifold gives a compact model, which we verify against the third-order Stokes wave and the Dysthe limit. The paper closes with numerical comparisons against the conformal-mapping reference, the cubic $(\eta,\phi)$ truncation and the super compact equation, for narrow-band (Benjamin--Feir) and broadband initial conditions.

\section{Governing equations}\label{sec:equations}

We consider two-dimensional, irrotational (potential) motion of an incompressible, inviscid, infinitely deep fluid with a free surface, subject to a constant gravitational field with acceleration $g$ acting vertically downwards. Throughout this analysis, we employ dimensionless variables and perturbation theory with respect to the small steepness parameter $\epsilon=a/l$, where $a$ is the characteristic wave amplitude and $l$ is the characteristic wavelength. The space-time variables are scaled as $x=\tilde{x}/l$, and $t=\sqrt{g/l}\,\tilde{t}$, where dimensional variables are denoted by a tilde.

The state of the fluid is described by the free-surface elevation 
$\eta(x,t)$ and the velocity potential at the free surface $\phi(x,t)$, which form canonically conjugate variables 
\citep{Zakharov1968}. These are scaled as $\eta=\tilde{\eta}/a$ and $\phi=\tilde{\phi}/(a\sqrt{lg})$. The evolution of the system is governed by Hamilton's equations,
\begin{equation} \label{eq:hamilton_laws}
\eta_t = \frac{\delta \mathcal{H}}{\delta \phi}, \quad \phi_t = -\frac{\delta \mathcal{H}}{\delta \eta},
\end{equation}
where $\delta/\delta$ denotes the variational derivative.  
The Hamiltonian $\mathcal{H}$ admits a systematic expansion in wave steepness, corresponding to the Taylor expansion of the Dirichlet-to-Neumann operator \citep{CraigSulem1993, DyachenkoKachulinZakharov2017}:
\begin{equation}
\label{eq:H_expansion}
\mathcal{H}=\mathcal{H}_2+\epsilon\mathcal{H}_3+\epsilon^2\mathcal{H}_4+\mathcal{O}(\epsilon^3)
\end{equation}
where the terms at each order in $\epsilon$ are
\begin{align*}                  
    \mathcal{H}_2&=\frac{1}{2}\int(\eta^2+\phi H\phi_x)\,\mathrm{d}x,\\
    \mathcal{H}_3&=\frac{1}{2}\int \left[(\phi_x)^2-(H\phi_x)^2\right]\eta\,\mathrm{d}x=\int \phi_xH\phi_x H\eta\,\mathrm{d}x,\\
    \mathcal{H}_4&=\frac{1}{2}\int \left[\eta^2\phi_{xx}H\phi_x+\phi H(\eta H(\eta H\phi_x)_x)_x\right]\mathrm{d}x.
\end{align*}
Here $H$ denotes the Hilbert transform with the Fourier symbol $-\ii\sgn(k)$, subscripts denote partial derivatives, and integrals are over the real line.

Retaining terms through $\mathcal{H}_4$ is necessary and sufficient for the weakly nonlinear dynamics of deep-water waves. Quadratic nonlinearity alone (generated by $\mathcal{H}_3$) produces no net self-modulation, since the deep-water dispersion relation $\omega^2 = |k|$ forbids three-wave resonances \citep{Phillips1960}. Cubic nonlinearity is the leading source of phenomena such as the Benjamin--Feir instability \citep{Benjamin_Feir_1967}, the Stokes-wave frequency shift, and the induced mean-flow coupling \citep{Dysthe1979}. Higher-order corrections contribute negligibly at moderate steepness, and the $\mathcal{H}_4$-truncated Hamiltonian model therefore agrees closely with full-Euler simulations for one-dimensional deep-water dynamics \citep{YUEN198267, WestEtAl1987, DyachenkoZakharov2011, Dyachenko2016_book}. 

The Hamiltonian \eqref{eq:H_expansion} yields, after some simplifications, the following equations of motion:
\begin{subequations} \label{eq:evolution_system}
\begin{align}
    \eta_t &= \begin{aligned}[t] 
        & H\phi_x - \epsilon(\eta\phi_x)_x - \epsilon H(\eta H\phi_x)_x \\
        & + \frac{\epsilon^2}{2}(\eta^2H\phi_x)_{xx} + \frac{\epsilon^2}{2}H(\eta^2\phi_{xx})_x + \epsilon^2H[\eta H (\eta H\phi_x)_x]_x + \mathcal{O}(\epsilon^3),
    \end{aligned} \label{eq:eta} \\[10pt]
    \phi_t &= \begin{aligned}[t]
        & -\eta + \epsilon H(\phi_x H\phi_x) \\
        & - \epsilon^2 H\phi_x \left[ \eta\phi_{xx} + H(\eta H\phi_x)_x \right] + \mathcal{O}(\epsilon^3).
    \end{aligned} \label{eq:phi}
\end{align}
\end{subequations}

Eliminating $\phi$ from these equations and retaining terms through $\mathcal{O}(\epsilon^2)$ yields the bidirectional equation for the surface elevation alone:
\begin{multline} \label{Matsuno_full}\eta_{tt}+H\eta_x=\epsilon\left[\eta\eta_x+H(\eta H\eta_x)-H((H\eta_t)^2)\right]_x\\
    +\epsilon^2\big\{2H(\eta_x\eta_t H\eta_t)-2\eta_t H(\eta H\eta_t)_x-\eta_x(H\eta_t)^2+H(\eta H[(\eta_t)^2]_x)\\
    -\tfrac{1}{2}(\eta^2H\eta_x)_x
-\tfrac{1}{2}H(\eta^2\eta_{xx})
-H(\eta H(\eta H\eta_x)_x)\big\}_x+\mathcal{O}(\epsilon^3).
  \end{multline}
We refer to this equation as the cubic deep-water Matsuno equation, first derived by \citet{Matsuno1993}.\footnote{The cubic-order derivation was not detailed in the original publications; we have independently verified \eqref{Matsuno_full}. Matsuno's convention for the Hilbert transform differs from ours by an overall sign.} This single-equation model can be considered as the deep-water analogue of the shallow-water Boussinesq equation (although the latter contains only quadratic nonlinearities). Despite its relevance, analytical studies on either the system \eqref{eq:evolution_system} or the equation \eqref{Matsuno_full} are lacking. This can likely be attributed to the intricate structure of the nonlinear terms and the prevalence of Fourier-space methods in deep-water wave theory.

\section{Canonical transformation}\label{sec:results}

We now perform a canonical change of variables that simplifies the Hamiltonian \eqref{eq:H_expansion} and the equations of motion \eqref{eq:evolution_system}. The transformation is motivated by the algebraic identity
\begin{multline}\label{eq:identity}
     -(\partial_{tt}+H\partial_x)\, H\!\left[\tfrac{1}{2}(H\eta)^2\right]=\\
     = \eta\eta_x + H(\eta H\eta_x) - H((H\eta_t)^2) 
     - H\!\left[H\eta\, H(\eta_{tt}+H\eta_x)\right],
\end{multline}
valid for any sufficiently smooth $\eta$. The first three terms on the right are exactly the quadratic nonlinearities of the Matsuno equation \eqref{Matsuno_full}; the last term contains the linear bidirectional operator $\partial_{tt} + H\partial_x$ and is of order $\mathcal{O}(\epsilon)$ on solutions. Consequently, Matsuno's equation at quadratic order factors as
\begin{equation}\label{Matsuno_factores_1}
    (\partial_{tt}+H\partial_x)\!\left[\eta+\epsilon\, H(H\eta\, H\eta_x)\right]
    =\mathcal{O}(\epsilon^2),
\end{equation}
revealing it as a linear equation up to a near-identity change of variables. This suggests defining a new variable $u$ via
\begin{equation}\label{eq:transform_O_eps}
    u = \eta + \epsilon\, H(H\eta\, H\eta_x) - \frac{\epsilon^2}{2} F(\eta).
\end{equation}
We fix $F(\eta)$ by requiring $\int \eta^2\, \mathrm{d}x = \int u^2\, \mathrm{d}x + \mathcal{O}(\epsilon^3)$.
This choice eliminates pure-$u$ terms in the new Hamiltonian at quartic order that would otherwise complicate the equations of motion. A short calculation yields 
\begin{equation}F(\eta) = H((H\eta)^2 H\eta_x)_x\equiv \frac{1}{3}H\left[(H\eta)^3\right]_{xx}.
\end{equation}
The corresponding transformation of $\phi$ is constructed via the generating functional $G[\eta, v] = \int v\, u(\eta)\, \mathrm{d}x$, with $u=\delta G/\delta v$ and $\phi = \delta G / \delta \eta$. Being of the mixed type, with old coordinate and new momentum as arguments, $G$ is the field-theoretic analogue of the classical generating function of the second kind, $F_2(q,P)$, and yields a canonical pair for any choice of $u(\eta)$. Computing the variational derivative $\phi = \delta G / \delta \eta$ gives $\phi$ as a functional of $v$ and $\eta$. Together with \eqref{eq:transform_O_eps} we have the mixed near-identity transformation
\begin{align}\label{transform_u(eta)}
    u =& \eta + \epsilon\, H(H\eta\, H\eta_x)
       - \frac{\epsilon^2}{2}\, H((H\eta)^2 H\eta_x)_x,\\
    \phi =& v - \epsilon\, H(H\eta\, Hv_x)
         - \frac{\epsilon^2}{2}\, H((H\eta)^2 Hv_{xx}),
\end{align}
which is exactly canonical by construction. Inverting order by order in $\epsilon$ gives, after simplifications, the explicit form
\begin{align}
\label{eq:can_eta}
    \eta =& u - \epsilon\, H(Hu\, Hu_x)
         - \frac{\epsilon^2}{2}\, H((Hu)^2 Hu_x)_x + \mathcal{O}(\epsilon^3),\\\label{eq:can_phi}
    \phi =& v - \epsilon\, H(Hu\, Hv_x)
         - \frac{\epsilon^2}{2}\, H((Hu)^2 Hv_x)_x + \mathcal{O}(\epsilon^3),
\end{align}
canonical through $\mathcal{O}(\epsilon^2)$. Equation \eqref{eq:can_eta} will later be used to recover the physical surface elevation $\eta$. In the new variables, the Hamiltonian reads
\begin{multline}\label{hamiltonian_uv}
\mathcal{H}=\frac{1}{2}\int (u^2+vHv_x)\,\mathrm{d}x\\
-\epsilon^2\int\left(v_{xx}H(uHu)+\frac{1}{2}HuH(HuHv_x)_x-\frac{1}{2}uH(uHv_x)_x\right)Hv_x\,\mathrm{d}x\\+\mathcal{O}(\epsilon^3).
\end{multline}
The canonical transformation has eliminated the cubic Hamiltonian $\mathcal{H}_3$, leaving only quadratic and quartic contributions at the order retained. Such transformations are standard in the Fourier-space Hamiltonian theory of weakly nonlinear surface waves, here performed in real variables. We note that the explicit form (\ref{eq:can_eta})--(\ref{eq:can_phi})
coincides through $\mathcal{O}(\epsilon^{2})$ with the Taylor expansion of the Hamiltonian flow by which
\citet{CraigGuyenneSulem2021} also eliminate $\mathcal{H}_3$ in real variables.

The equations of motion now take the form
\begin{subequations}\label{eq:uv_system}
\begin{align}
\label{full_uv_1}
    u_t=\frac{\delta \mathcal{H}}{\delta v}=&H\partial_x(v+\epsilon^2 p)+\mathcal{O}(\epsilon^3),\\[5pt]
\label{full_uv_2}
    v_t=-\frac{\delta \mathcal{H}}{\delta u}=&-u+\epsilon^2 q+\mathcal{O}(\epsilon^3),
\end{align}
\end{subequations}
where, after simplifications using the Tricomi identity \citep{King2009Vol1} and standard differentiation rules, the cubic terms read
\begin{subequations} \label{eq:pq_of_uv_system}
\begin{align}
    p &= \begin{aligned}[t]
        & H(uv_xHu)_x-HuH(uv_x)_x \\
        & -uH(v_xHu)_x+uHuHv_{xx},
    \end{aligned} \label{eq:eq_p} \\[5pt]
    q &= \begin{aligned}[t]
        & -H(v_xHuHv_x)_x+v_xH(HuHv_x)_x \\
        & +u_xv_xHv_x+Hv_xH(v_xHu)_x.
    \end{aligned} \label{eq:eq_q}
\end{align}
\end{subequations} 
Notably, evaluating $p$ and $q$  under a monochromatic ansatz $u = a\cos(kx)$, $v = b\sin(kx)$ reveals that all forced harmonics at wavenumber $3k$ cancel identically:
\begin{equation}
    p=\frac{1}{2}a^2bk^2\sin(kx)=\frac{1}{2}a^2k^2v,\quad q=-\frac{1}{2}ab^2k^3\cos(kx)=-\frac{1}{2}b^2k^3u.
\end{equation}
Individually, each of the eight cubic terms always produces a $3k$ harmonic. This is therefore an exact structural cancellation. The canonical transformation ensures that the second- and third-harmonic forcing vanishes on the wavenumber diagonal. Consequently, the third-order Stokes wave collapses to a pure 
fundamental harmonic in $(u,v)$ space (see also \S\ref{sec:Stokes} and \S\ref{sec:Dysthe}). The only 
surviving cubic effect in this case is the resonant four-wave self-interaction responsible for the amplitude-dependent frequency shift. While all bound harmonics of the Stokes wave are eliminated from $(u,v)$, they are fully recovered through the back-transformation \eqref{eq:can_eta}, \eqref{eq:can_phi}.

The bidirectional Hamiltonian system \eqref{eq:uv_system} can be viewed as a real-variable counterpart to the $(1+1)$-dimensional Krasitskii-reduced Zakharov equation \citep{Zakharov1968, Krasitskii1994}, or to the canonical system for counter-propagating waves introduced by \citet{Dyachenko2020canonical}, who splits the complex normal variable into right- and left-moving parts and treats the two groups as separate fields. These three formulations use distinct representations, but are canonically equivalent and govern identical dynamics at this order. The present one keeps the fields real and close to the physical variables, with $\eta$ and $\phi$ recovered by a relatively simple map. When expressed in normal variables, the Fourier kernel of \eqref{hamiltonian_uv} agrees with Krasitskii's four-wave kernel on the resonant manifold; off resonance, it realizes a gauge-equivalent form \citep[see also][]{DyachenkoKachulinZakharov2017}. The closed form of the kernel is given in appendix~\ref{appA}.

For completeness, the mass, momentum, and energy conservation laws in the canonical variables $(u,v)$ are, respectively,
\begin{equation}\label{conservation_laws}
    \int u\, \mathrm{d}x=\mathrm{const}, \quad \int uv_x\, \mathrm{d}x =\mathrm{const}, \quad \mathcal{H}=\mathrm{const}.
\end{equation}

\section{Unidirectional reduction}\label{sec:unidirectional_reduction}

\subsection{Linear reduction}
To restrict the system to unidirectional waves, we start by considering only the linear terms of \eqref{eq:uv_system}, which are now correct up to $\mathcal{O}(\epsilon)$. Eliminating $v$ yields the linear Hilbert wave equation for $u$:
\begin{equation}
\label{u_tt_linear}
    u_{tt}+Hu_x= \mathcal{O}(\epsilon^2).
\end{equation}
This is just the quadratic Matsuno equation \eqref{Matsuno_factores_1} in terms of $u$.

We introduce the pseudo-differential operator $A$ with Fourier symbol $\ii\sgn(k)\sqrt{|k|}$, satisfying
\begin{equation}\label{eq:Asquared}
A^2 = -H\partial_x,\quad A^{-1}=AH\partial_x^{-1}.
\end{equation}
This operator factors the linear bidirectional equation \eqref{u_tt_linear} as $(\partial_t - A)(\partial_t + A)u = \mathcal{O}(\epsilon^2)$. Selecting the right-going factor yields the unidirectional equation
\begin{equation}
\label{lin_Au}
    u_t + Au = \mathcal{O}(\epsilon^2)
\end{equation}
with dispersion relation $\omega = \sgn(k)\sqrt{|k|}$. The phase velocity $\omega/k = 1/\sqrt{|k|}$ is positive for all $k \neq 0$, ensuring that all Fourier components propagate to the right. This is the real-variable counterpart of the complex-valued formulation in which unidirectionality is imposed by restricting the spectrum to $k > 0$ with $\omega = \sqrt{k}$ \citep[see for example][]{DyachenkoKachulinZakharov2017}. In the present real-valued setting, both positive and negative wavenumbers are retained, and the directional selection is encoded entirely in the dispersion relation.

Combining \eqref{lin_Au} with the linear part of equation~\eqref{full_uv_1} gives the unidirectional reduction
\begin{equation}\label{v(u)_linear}
    v = A^{-1} u+ \mathcal{O}(\epsilon^2),
\end{equation}
fully consistent with the Hamiltonian formulation through $\mathcal{O}(\epsilon)$. The reduced equation \eqref{lin_Au} can be written as a standard Hamiltonian PDE: 
\begin{equation}
    u_t=J\left(\frac{\delta\mathcal{H}_{\mathrm{red}}}{\delta u}\right)
\end{equation}
with the Poisson operator $J=-\tfrac{1}{2}A$ and the Hamiltonian $\mathcal{H}_{\mathrm{red}}=\int u^2 \mathrm{d}x+\mathcal{O}(\epsilon^2)$.

\subsection{Reduction up to $\mathcal{O}(\epsilon^2)$}

In the present work, we do not pursue the exact canonical reduction at this order in
real-space variables. The canonical route is fully realized by the super compact equation using Fourier-space methods \citep{DyachenkoKachulinZakharov2017}: a carefully constructed canonical transformation absorbs most of the non-resonant content and complexity into the variable definition, reducing the four-wave dynamics to just two nonlinear terms. The corresponding transformation in real-space variables does not seem to admit a closed-form expression in elementary operators ($\partial_x$, $H$, $A$). We instead retain the variable $u$, which remains closer to the physical surface elevation $\eta$ through the explicit transformation~\eqref{eq:can_eta}, and accept a non-canonical reduction at $\mathcal{O}(\epsilon^2)$.

We seek an evolution equation in the form
\begin{equation}\label{eqn:ut_formal}
    u_t  +A u = \epsilon^2 H \partial_x f+ \mathcal{O}(\epsilon^3),
\end{equation}
where $f(u)$ is a cubic functional to be determined. Substituting \eqref{v(u)_linear} and \eqref{eqn:ut_formal} into the system \eqref{eq:uv_system} yields the reduction
\begin{equation}\label{reduction}
    v = A^{-1} u + \epsilon^2 (f - p) + \mathcal{O}(\epsilon^3),
\end{equation}
subject to the compatibility condition
\begin{equation}\label{reduct_1}
    f_t - A f = q + p_t\equiv r.
\end{equation}
Equation \eqref{reduct_1} is a non-local functional equation for $f$. To evaluate the right-hand side $r$, we use the linear relations $v_x = -Hu_t + \mathcal{O}(\epsilon^2)$ and $v_{xx} = -Hu_{xt} + \mathcal{O}(\epsilon^2)$ to eliminate $v$. Because $A$ does not satisfy a Leibniz rule, deferring its substitution until the final step makes the algebraic manipulation of $r$ substantially more tractable. Using the Tricomi identity $H(f_1f_2) = H(Hf_1\,Hf_2) + f_1\,Hf_2 + f_2\,Hf_1$, standard derivative rules, and $u_{tt} = -Hu_x + \mathcal{O}(\epsilon^2)$ to eliminate second-order time derivatives, the source $r$ is expressed in terms of $u$ and $u_t$:
\begin{multline}\label{q+pt_1}
    r=-u_{x}H(u_{t}^{2} + u_{x}Hu) - H[Hu\, H(u_{t}^{2} + u_{x}Hu)]_{x}\\
    +\,2H(u_t Hu\, Hu_t)_x - 2 Hu_{t}H(u_{t}Hu)_{x} + \tfrac{1}{2}Hu\,H\!\left[(Hu)^2\right]_{xx}+\mathcal{O}(\epsilon^2).
\end{multline}
Substituting $u_t = -Au + \mathcal{O}(\epsilon^2)$ then yields the purely spatial unidirectional form:
\begin{multline}\label{q+pt_2}
    r=-u_{x}H((Au)^{2} + u_{x}Hu) - H\left[Hu\, H((Au)^{2} + u_{x}Hu)\right]_x\\
    +2H(Au\,Hu\,AHu)_x - 2 H(Au\,Hu)_{x}AHu + \tfrac{1}{2}Hu\,H\!\left[(Hu)^2\right]_{xx}+\mathcal{O}(\epsilon^2).
\end{multline}

Note that the bidirectional equation corresponding to Matsuno's equation \eqref{Matsuno_full} is 
\begin{equation}
    u_{tt}+Hu_x=\epsilon^2H\partial_x r+\mathcal{O}(\epsilon^3),
\end{equation}
with $r$ given by \eqref{q+pt_1}.
\subsection{Approximate closure}
To obtain the unidirectional reduction \eqref{reduction} explicitly, we need to solve the functional equation \eqref{reduct_1} for $f(u)$. However, an exact solution is not readily obtainable. We instead exploit a rewriting of \eqref{eqn:ut_formal} and \eqref{reduct_1} that exposes the structure for an approximate closure:
\begin{equation}\label{new1}
    u_t + Au = \frac{\epsilon^2}{2} A\bigl[r - (f_t + Af)\bigr]
    + \mathcal{O}(\epsilon^3),
\end{equation}
\begin{equation}\label{new2}
   f = -\frac{1}{2}A^{-1}\bigl[r - (f_t + Af)\bigr].
\end{equation}

The dynamics in \eqref{new1} depends on $f$ only through the dispersive combination $f_t + Af$. This structure suggests an iterative solution to \eqref{new2} beginning with the leading-order approximation
\begin{equation}\label{closure}
    f_t + Af = 0.
\end{equation}
While successive iterations would systematically correct the residual at each step, we do not pursue them here. Instead, we adopt \eqref{closure} as a natural unidirectional ansatz. The operator $\partial_t + A$ governs the rightward-propagating branch of the linear dynamics; thus, \eqref{closure} constrains $f$ to the same dispersion shell on which $u$ already lies by construction. This closure is exact whenever the nonlinear source $r$ is itself concentrated on this shell. More specifically, imposing \eqref{closure} introduces an error of order $\epsilon^2(r_t+Ar)$ in the evolution equation \eqref{new1}. This error, however, vanishes precisely where the leading nonlinear interactions are most significant, which we now demonstrate.

A direct calculation of the Fourier symbol of $r$ shows that its symmetric kernel, $M^{\mathrm{sym}}(k_1, k_2, k_3)$, vanishes unless the wavenumbers satisfy $\sgn(k_1) + \sgn(k_2) + \sgn(k_3) = \sgn(k)$, where $k = k_1+k_2+k_3$ is the resultant wavenumber (see appendix \ref{app:kernel} for an exact formula for the kernel). This condition restricts the spectral support of the cubic source $r$ to the $(++-)$ configuration for $k > 0$ and the $(--+)$ configuration for $k < 0$. On this support, the resonance condition $\Delta\omega\equiv\omega(k_1)+\omega(k_2)+\omega(k_3)-\omega(k)=0$, with
$\omega(k)=\sgn(k)\sqrt{|k|}$, admits only trivial solutions, in which the wavenumbers pair off as $\{k_1,k_2\}=\{k,-k_3\}$. At leading order, each factor of $u$ in $r$ evolves freely, $\hat{u}(k_i)\propto\mathrm{e}^{-\ii\,\omega(k_i)t}$, and the integrand of $\hat{r}$ oscillates at $\omega(k_1)+\omega(k_2)+\omega(k_3)$. The operator
$\partial_t+A$ multiplies it by $-\ii\,\Delta\omega$, leading to the relation $r_t + Ar = 0$ on the resonant manifold. Consequently, the closure \eqref{closure} is exact on the resonant manifold and approximate elsewhere, with a residual that scales as the resonance defect $\Delta\omega$ and vanishes as $\mathcal{O}(\delta^2)$ for a narrow-band field $u$ of relative bandwidth $\delta$. When expressed in normal variables, the kernel agrees on the resonant manifold with the four-wave coefficient of the Zakharov equation \citep{Zakharov1968,Krasitskii1994,DyachenkoKachulinZakharov2017}.

With \eqref{closure}, the system closes: \eqref{new2} gives $f=-\tfrac{1}{2}A^{-1}r$, and \eqref{new1} reduces to the unidirectional evolution equation
\begin{equation}\label{eq:full_reduction}
    u_t + Au = \frac{\epsilon^2}{2}A r,
\end{equation}
with $r$ as in \eqref{q+pt_2}. 

Equation~\eqref {eq:full_reduction} is the central evolution equation of this paper. It shares its resonant four-wave
content with the compact and super compact equations of \citet{DyachenkoZakharov2011,DyachenkoZakharov2012} and
\citet{DyachenkoKachulinZakharov2017}, differing from both effectively by a linear change of variables and a kernel term that vanishes on the resonant manifold. Unlike those complex-variable formulations, \eqref{eq:full_reduction} is
not Hamiltonian at $\mathcal{O}(\epsilon^2)$: its four-wave kernel (appendix~\ref{app:kernel}) lacks the wavenumber symmetry that a Hamiltonian origin would impose \citep[see also][]{Krasitskii1994}. The practical cost is mostly structural rather than dynamical. Mass conservation $\int u\,\mathrm{d}x$ remains exact, while energy and momentum in~\eqref{conservation_laws}, with $v$ slaved as in~\eqref{reduction}, are conserved approximately, with small, bounded $\mathcal{O}(\epsilon^2)$ relative oscillations and no secular drift at this order. Hamiltonian tools, such as the Noether correspondence between symmetries and conservation laws, or variational stability arguments for coherent structures (breathers), are lost. Nevertheless, the kernel of \eqref{eq:full_reduction} coincides with that of the Hamiltonian relatives on the resonant manifold, and the close agreement between \eqref{eq:full_reduction} and the super compact equation in \S\ref{sec:Numerics} is a direct consequence of this structure.

\section{Simplified model equation} \label{sec:simplified_model}

A further simplification of equation \eqref{eq:full_reduction} yields a more compact model. The first two terms of \eqref{q+pt_2} share the factor $Q(u) = (Au)^2 + u_x Hu$ with Fourier symbol $\tfrac{1}{2}\sgn(k_1 k_2)(\sqrt{|k_1|} -\sqrt{|k_2|})^2$, which vanishes quadratically on the resonant diagonal $k_1 = k_2$. On monochromatic input, $Q(u)$ vanishes identically; on narrow-band wavepackets, $Q(u) = \mathcal{O}(\delta^2)$ for bandwidth $\delta$. Notably, the canonical transformation \eqref{eq:can_eta} contributes bound harmonics to $\eta$, so a field narrow-band in $u$ generally appears multi-modal in $\eta$.

Neglecting the $Q$-terms and rewriting the remainder in commutator
form, we propose the reduced model
\begin{equation}\label{main_eqn}
    u_t+Au=\epsilon^2 A\left\{[H\partial_x, AHu](AuHu)
    + \tfrac{1}{4}HuH[(Hu)^2]_{xx}\right\},
\end{equation}
as an approximation of \eqref{eq:full_reduction}, where $[H\partial_x, AHu]\psi = H(AHu\,\psi)_x - H(\psi)_x AHu$.
The dimensional form is obtained by setting $\epsilon=1$ and inserting $\sqrt{g}$ in front of the two outer $A$-operators.

Equation \eqref{main_eqn} retains the full linear dispersion and captures the bulk of the resonant cubic interactions of the parent system across both narrow-band and broadband initial conditions (see \S\ref{sec:Numerics}). The neglected $Q$-terms carry the $\mathcal{O}(\delta^2)$ sub-leading dynamics. They become essential only when nonlinear focusing drives energy into a high-wavenumber spectral tail at fine numerical resolution, providing the high-wavenumber regularization that \eqref{main_eqn} otherwise lacks. We return to this point in the next section (\S\ref{sec:stability}). Because $Q(u)$ vanishes on monochromatic input and enters at $\mathcal{O}(\epsilon^4)$ under the multiple-scales expansion, the following results hold equally for \eqref{eq:full_reduction}.

\subsection{Stokes wave}\label{sec:Stokes}
As a consistency check, equation \eqref{main_eqn} admits the exact
monochromatic travelling-wave solution
\begin{equation}
    u = C\cos\theta, \qquad \theta = k_0 x - \omega t,
    \qquad \omega = \sqrt{k_0}\left(1 + \tfrac{1}{2} \epsilon^2C^2 k_0^2\right),
\end{equation}
with arbitrary $k_0>0$ and $C$. (The equation is also invariant under constant shifts $u \mapsto u + C_0$, since $A$, $H$, and $r$ all annihilate constants.) This is verified directly using
\begin{align*}
H[\sin(k_0x)] &= -\cos(k_0x), & H[\cos(k_0x)] &= \sin(k_0x),\\
A[\sin(k_0x)] &= \sqrt{k_0}\cos(k_0x), & A[\cos(k_0x)] &= -\sqrt{k_0}\sin(k_0x),
\end{align*}
together with standard trigonometric identities. Choosing $C = a\bigl(1 + \tfrac{\epsilon^2}{8} a^2 k_0^2\bigr)$, so that the first-harmonic coefficient of $\eta$ equals $a$, the canonical transformation \eqref{eq:can_eta} gives
\begin{align}
    \eta &= a\cos\theta + \frac{1}{2} \epsilon k_0a^2 \cos 2\theta 
         + \frac{3}{8} \epsilon^2k_0^2a^3  \cos 3\theta + \mathcal{O}(\epsilon^3),\\
    \omega &= \sqrt{k_0}\left(1 + \frac{1}{2} \epsilon^2k_0^2a^2 \right) 
         + \mathcal{O}(\epsilon^3).
\end{align}
This is exactly the classical third-order Stokes solution for deep-water gravity waves, in agreement with the result obtained from the bidirectional cubic equation \eqref{Matsuno_full} by \citet{Matsuno1993}.

\subsection{Recovery of the Dysthe equation}\label{sec:Dysthe}
We now derive the envelope equation for slowly modulated wavepackets governed by \eqref{main_eqn}. We introduce the standard multiple-scales ansatz for a narrow-band wavepacket centred at carrier wavenumber $k_0$:
\begin{equation}\label{ansatz}
    u(x, t) = \Psi(\xi, \tau)\, \mathrm{e}^{\ii\theta} +\epsilon\,\Psi_2(\xi, \tau)\, \mathrm{e}^{2\ii\theta}+\epsilon^2\,\Psi_3(\xi, \tau)\, \mathrm{e}^{3\ii\theta}+ \text{c.c.} 
    + \mathcal{O}(\epsilon^3),
\end{equation}
with $\theta = k_0 x - \omega_0 t$, slow space $\xi = \epsilon\cdot(x - v_g\, t)$, and slow time $\tau = \epsilon^2 t$. Since slow spatial variation corresponds to narrow support in wavenumber $k$, the parameter $\epsilon$ now serves a twofold role, measuring both wave steepness and spectral bandwidth, and setting the NLS regime where cubic nonlinearity and group-velocity dispersion balance ($\delta=\epsilon$). The Dysthe terms then appear at $\mathcal{O}(\epsilon^3)$: the third-order dispersion, of order $\delta^3$, and the bandwidth corrections to the cubic nonlinearity, of order $\epsilon^2\delta$, both within the $\mathcal{O}(\epsilon^2)$ steepness truncation of the model. Formally, the same nominal order also reaches $\epsilon^3\mathcal{H}_5$ and a fourth harmonic $\epsilon^3\Psi_4\mathrm{e}^{4\ii\theta}+\mathrm{c.c.}$, both third order in steepness and outside the truncation, but neither contributes to the envelope equation at $\mathcal{O}(\epsilon^3)$ even if kept.

For a product of functions with non-overlapping Fourier spectra and $|\kappa|\gg|k|$ we have $\sgn(k+\kappa)=\sgn(\kappa)$ and the Fourier symbol of $A$   at total wavenumber $k+\kappa$ admits the expansion
\begin{multline} K_{A}(k, \kappa)=\ii\sgn(k+\kappa)\sqrt{|k+\kappa|}\\ =\ii\sgn{(\kappa)}\sqrt{|\kappa|}+\frac{1}{2\sqrt{|\kappa|}}\ii k+\frac{\ii\sgn(\kappa)}{8|\kappa|^{3/2}}(\ii k)^2
    -\frac{1}{16 |\kappa|^{5/2}}(\ii k)^3+...
\end{multline}
Therefore
\begin{equation}
    A(\Psi \mathrm{e}^{\ii\theta})=\left(\ii\sqrt{k_0}\Psi+\frac{\epsilon}{2\sqrt{k_0}}\Psi_\xi+\frac{\ii\epsilon^2}{8k_0^{3/2}}\Psi_{\xi\xi}-\frac{\epsilon^3}{16 k_0^{5/2}}\Psi_{\xi\xi\xi}\right)\mathrm{e}^{\ii\theta}+\mathcal{O}(\epsilon^4),
\end{equation}
with analogous expansions for complex-conjugate components and other products in nonlinear terms.

Substituting~\eqref{ansatz} into~\eqref{main_eqn} and matching orders in $\epsilon$ at the carrier $\mathrm{e}^{\ii\theta}$ produces the standard hierarchy: orders $\mathcal{O}(\epsilon^0)$ and $\mathcal{O}(\epsilon^1)$ give $\omega_0=\sqrt{k_0}$ and $v_g=1/(2\sqrt{k_0})$; order $\mathcal{O}(\epsilon^2)$ produces NLS; order $\mathcal{O}(\epsilon^3)$ produces corrections to NLS. Importantly, isolating the coefficients of $\mathrm{e}^{2\mathrm{i}\theta}$ and $\mathrm{e}^{3\mathrm{i}\theta}$ yields $\Psi_2=\Psi_3=0$. Thus, for slowly modulated wavepackets, the bound second and third harmonics do not enter the dynamics of the canonical variable $u$; as for the Stokes wave, they are completely absorbed into the transformation \eqref{eq:can_eta}. 

Finally, combining terms at the carrier $\mathrm{e}^{\ii\theta}$ through order $\mathcal{O}(\epsilon^3)$ yields the envelope equation
\begin{multline}\label{eq:dysthe}
    \Psi_\tau + \frac{\ii}{8 k_0^{3/2}}\Psi_{\xi\xi} 
    - \frac{\epsilon}{16 k_0^{5/2}}\Psi_{\xi\xi\xi}
    = -2\ii k_0^{5/2}|\Psi|^2 \Psi \\
    + \epsilon\, k_0^{3/2}\big(-6|\Psi|^2\Psi_\xi - \Psi^2\bar\Psi_\xi 
    + 2\ii\,\Psi\, H(|\Psi|^2)_\xi\big),
\end{multline}
which is the $(1+1)$-dimensional Dysthe equation in its 
original form \citep{Dysthe1979,Janssen_1983,STIASSNIE1984}, distinct from later Hamiltonian variants. The left-hand side contains the linear group-velocity-frame dispersion at second and third order, the right-hand side contains the leading cubic self-interaction (NLS), local frequency-dependent corrections, and the nonlocal coupling to the induced mean flow.

In the standard Dysthe derivations, the nonlocal term $\Psi H(\vert{}\Psi\vert{}^2)_\xi$ is obtained by introducing a mean velocity potential, solving its Laplace-type boundary-value problem, and matching the resulting mean flow to the envelope equation \citep{Dysthe1979, TrulsenDysthe1996}. This auxiliary step is bypassed in Zakharov's Hamiltonian description, where the mean-flow coupling is natively captured \citep{STIASSNIE1984, DyachenkoKachulinZakharov2017envelope}. The present formulation, being of the same origin, shares this structural efficiency. The mean-flow content appears here as the $\mathrm{e}^0$ harmonic of the combination $AuHu$ within the middle cubic $H(AuHu)_{x} AHu$ of \eqref{main_eqn}, is made nonlocal by the outer Hilbert transform, and is fed back to the carrier frequency $\mathrm{e}^{\ii\theta}$ via the right-most $AHu$ factor.

\subsection{A minimal model}\label{sec:minimal_model}

Examining the individual contributions of the three cubic terms in 
\eqref{main_eqn}, the middle cubic, $H(Au\, Hu)_x\, AHu$, alone 
reproduces the NLS self-interaction and the exact Dysthe nonlocal mean-flow term. Formally, retaining only this term yields a particularly simple model equation 
\begin{equation}
    \label{main_eqn_minimal}  u_t+Au+\epsilon^2\,A\left\{  H(AuHu)_{x} AHu \right\}=0,
\end{equation}
which combines the full linear dispersion with the dominant cubic 
self-interaction. The local self-steepening contributions of the remaining two cubics enter only with modified coefficients. In numerical simulations, this minimal model becomes inaccurate in regions of high local amplitude, but captures the 
dominant Dysthe physics. We emphasize that it is presented here as a structural observation rather than a proposed replacement for \eqref{main_eqn} and is not pursued further here.

\section{Numerical examples}\label{sec:Numerics}

The numerical simulations presented below are primarily confirmatory: we verify that the reduced model \eqref{main_eqn}, marked as M1 on the figures, behaves consistently with the original truncated $\eta$--$\phi$ system \eqref{eq:evolution_system}, the super compact equation \citep{DyachenkoKachulinZakharov2017}, and the fully nonlinear Euler equations in conformal variables \citep{DyachenkoEtAl1996}---the latter serving as a proxy for the exact bidirectional dynamics. We do not analyse relative accuracy. The three $\mathcal{O}(\epsilon^2)$ truncations ($\eta$--$\phi$, our model, and SC) share the same formal asymptotic ordering, and apparent discrepancies between them are dominated by other effects explained below, rather than intrinsic model fidelity. For visual clarity, only the simplified model \eqref{main_eqn} is plotted (M1). The full \eqref{eq:full_reduction} yields marginally closer agreement with the super compact equation, but the difference is not visually discernible.
 
The comparison with the super compact equation is subject to a small initial-condition (IC) mismatch at $\mathcal{O}(\epsilon^2)$. The super compact variable $c$ is related to the physical surface elevation $\eta$ by a near-identity transformation expanded to second order in $\epsilon$ in \citet{DyachenkoKachulinZakharov2017}; to our knowledge, the third-order term has not been published. Consequently, the transformation from a prescribed $\eta$ profile to the corresponding $c$ profile (and back) is accurate only to $\mathcal{O}(\epsilon)$, leaving an unavoidable $\mathcal{O}(\epsilon^2)$ mismatch in the initial wave profile. Because the super compact equation is unidirectional by construction, this surface mismatch propagates within the rightward-going dynamics without generating leftward radiation, keeping its overall effect on the comparison small.

The comparison with the full Euler equations is necessarily approximate. An initial condition prescribed as a rightward-going profile of \eqref{main_eqn}, when projected into the bidirectional Euler system, generates a small amount of leftward radiation. This radiation carries away a portion of the energy and momentum that would otherwise belong to the rightward content, effectively reshaping the rightward-going initial condition. Under the periodic boundary conditions used in the integrations, the leftward radiation also remains in the computational domain and slightly distorts the rightward dynamics throughout the simulation.

Solver details are presented in appendix \ref{sec:app_numerics}.

\subsection{Broadband initial condition}

The first test exercises the four models on a broadband initial condition with substantial spectral spread, operating in a regime where the structural simplifications of \eqref{main_eqn} are no longer formally justified. We take
\begin{equation}
    \eta(x,0)=a\cos{\left(b\,[x+\cos{(cx-x_1)}]\right)}\,\mathrm{e}^{-d(x-x_2)^2}
\end{equation}
with $a=1.95\times10^{-3}$, $b=25$, $c=1.5$, $d=1.6$, $x_1=0.5$, $x_2=\pi-0.4$. This packet has spectral peak at $k_{\mathrm{peak}} \approx 59$ and spectral mean $\langle k \rangle \approx 40$, with normalised spectral width $\Delta k / k_{\mathrm{peak}} \approx 1.14$, where $\Delta k$ is the wave number range over which $|\hat\eta(k)|^2$ exceeds $1\%$ of its peak value. The initial maximum steepness is $\max|\eta_x| \approx 0.09$, growing up to $0.31$ during the focusing phase near $t = 33$ before defocusing to a modulated wave-train.

\begin{figure}
  \centerline{\includegraphics{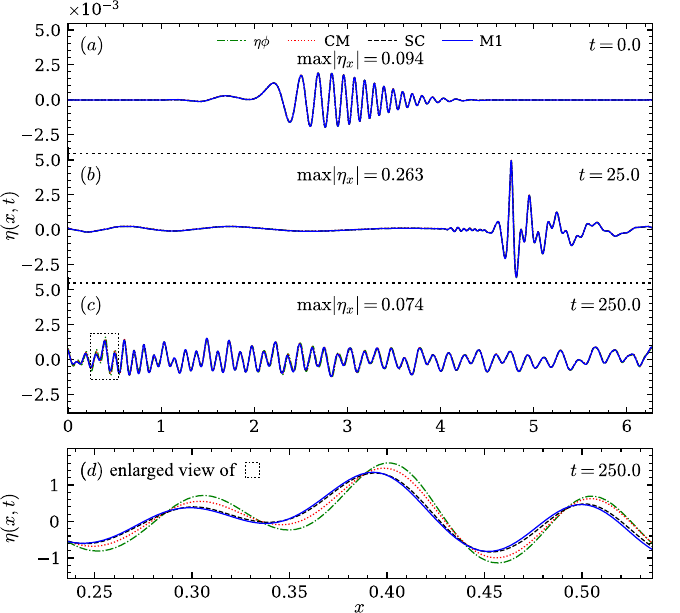}}
  \caption{Free-surface elevation $\eta(x,t)$ for the broadband initial condition. Evolved with the conformal-mapping reference (CM, red dotted), the third-order $\eta$--$\phi$ truncation ($\eta\phi$, green dash-dotted), the super compact equation (SC, black dashed), and the unidirectional model \eqref{main_eqn}, named as M1 (blue
  solid). (\textit{a}) Initial condition, $\max|\eta_x| = 0.09$. (\textit{b}) State at $t = 25$, during focusing, $\max|\eta_x| = 0.26$. The global peak steepness of $0.31$ is attained near $t = 33$. (\textit{c}) Modulated wavetrain at $t = 250$ following defocusing, $\max|\eta_x| = 0.07$. (\textit{d}) Enlarged view of the boxed region in
  (\textit{c}), showing systematic phase drift at late time. All four curves overlap to plotting accuracy in (\textit{a}) and (\textit{b}). Domain $L = 2\pi$, $N = 2048$, $\Delta t = 5\times 10^{-3}$, $g = 1$ (nondimensional
  units throughout).}
\label{fig:broadband}
\end{figure}

\begin{figure}
  \centerline{\includegraphics{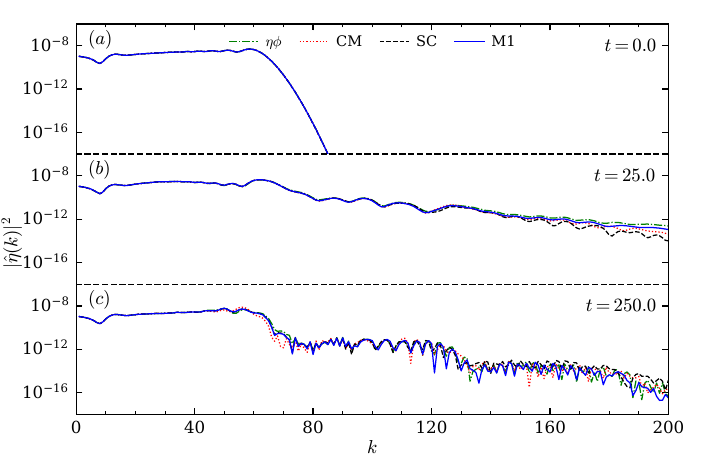}}
  \caption{Wavenumber spectra $|\hat\eta(k)|^2$ of the simulation of figure~\ref{fig:broadband}, at (\textit{a}) $t = 0$, (\textit{b}) $t = 25$, (\textit{c}) $t = 250$. Models and line styles as in figure~\ref{fig:broadband}. All plotted spectral content is significant; the double-precision rounding plateau lies many orders of magnitude below the plotted range.}
\label{fig:broadband_spectra}
\end{figure}

Figure~\ref{fig:broadband} shows the free-surface elevation at three representative times. At $t = 0$
(figure~\ref{fig:broadband}\textit{a}) and through the focusing event (figure~\ref{fig:broadband}\textit{b}, $t = 25$,
$\max|\eta_x| = 0.26$) all four models overlap on the plotting scale. By $t = 250$ (figure~\ref{fig:broadband}\textit{c}), the packet has dispersed into a long-wavelength modulated train spanning the periodic domain. The enlarged view (figure~\ref{fig:broadband}\textit{d}) shows the first systematic departures.
The corresponding wavenumber spectra are shown in
figure~\ref{fig:broadband_spectra}.

M1 \eqref{main_eqn} reproduces the broadband dynamics of the full-Euler reference at peak steepness $\max|\eta_x| \approx 0.31$, well into the regime where Dysthe's narrow-band ansatz no longer applies and approaching the breaking threshold $\max|\eta_x| \approx 0.4$ for deep-water surface waves. The agreement degrades somewhat for steeper initial conditions. For these cases, a higher spatial resolution may also be needed, motivating the additional regularising structure of \eqref{eq:full_reduction}, as explained in more detail in \S\ref{sec:stability}.

\subsection{Modulated wave-train initial condition}

We now take the initial condition as
\begin{equation}\label{eq:ic_bf}
\eta(x,0) = a\cos(k_0 x) + \tfrac{1}{2}k_0 a^2\cos(2k_0 x) 
+ \delta\bigl[\cos((k_0{+}p)x) + \cos((k_0{-}p)x)\bigr],
\end{equation}
with $a=2.5\times10^{-3}$, $k_0 = 32$, steepness $k_0 a = 0.08$, $p = 3$, and seed amplitude 
$\delta = 0.03\,a$. The Stokes bound harmonic is included to avoid 
spurious transients from the initial relaxation onto the Stokes branch. This is a standard configuration for the Benjamin--Feir modulational instability: a periodic second-order Stokes wave perturbed by a small sideband modulation well inside the instability band. The initial-condition spectrum is essentially monochromatic, placing this test, at least initially, inside the narrow-band regime where the structural simplification underlying M1 is theoretically justified
(\S\ref{sec:simplified_model}). The simulation is integrated to $T = 500$, well beyond the
first recurrence of the instability, on the domain $L = 2\pi$ with $N = 1024$
points and $\Delta t = 5\times 10^{-2}$.

\begin{figure}
  \centerline{\includegraphics{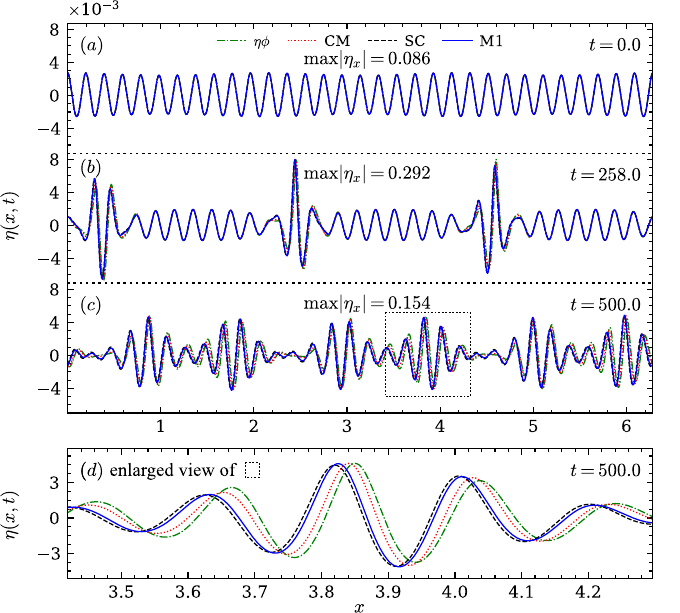}}
  \caption{Free-surface elevation $\eta(x,t)$ for the Benjamin--Feir initial condition \eqref{eq:ic_bf} with $a=2.5\times10^{-3}$, $k_0 = 32$, $k_0 a = 0.08$, $p = 3$, $\delta = 0.03\,a$. Models and line styles as in figure~\ref{fig:broadband}. (\textit{a}) Initial Stokes wavetrain with imperceptible sideband modulation, $\max|\eta_x| = 0.086$. (\textit{b}) Fully-developed modulational instability at $t = 258$, $\max|\eta_x| = 0.29$; three periodic envelope packets have emerged. (\textit{c}) Quasi-recurrent state at $t = 500$, $\max|\eta_x| = 0.15$. (\textit{d}) Enlarged view of the boxed region in (\textit{c}), showing the phase drift of $\eta\phi$ relative to CM at late time; M1 and SC remain close. Simulation parameters: $L = 2\pi$, $N = 1024$, $\Delta t = 5\times 10^{-2}$, $g = 1$.}
\label{fig:BF}
\end{figure}

Figure~\ref{fig:BF} shows the free-surface elevation at three representative times. The initial condition (figure~\ref{fig:BF}\textit{a}, $\max|\eta_x| = 0.086$) is a uniform Stokes wavetrain with imperceptible sideband perturbation. By $t = 258$ (figure~\ref{fig:BF}\textit{b}, $\max|\eta_x| = 0.29$) the instability has fully developed: three periodic envelope packets have emerged from the modulated train, each attaining a peak local steepness comparable to the breaking threshold for deep-water surface waves. By $t = 500$ (figure~\ref{fig:BF}\textit{c}, $\max|\eta_x| = 0.15$), the dynamics has entered a quasi-recurrent regime in which the envelope packets have broken up into a more complex modulated structure. The enlarged view (figure~\ref{fig:BF}\textit{d}) reveals the same model ordering observed in the previous broadband test.

The agreement of M1 with three independently derived references 
($\eta\phi$, SC, CM) over the BF growth phase and the first several recurrence cycles is the central numerical result of this section: four distinct models of the deep-water wave problem, prepared from the same initial condition through different projections and integrated with independent solvers, produce mutually consistent dynamics. Beyond this window, model-to-model differences are dominated by IC-projection mismatch, accumulated leftward radiation under periodic boundary conditions, and solver-specific effects rather than by intrinsic model accuracy; we therefore do not interpret late-time discrepancies as quantitative accuracy benchmarks.

\subsection{About numerical stability}\label{sec:stability}

Although the four $Q$-terms vanish on the trivial resonant manifold, and contribute only at $\mathcal{O}(\delta^2)$ on narrow-band wavepackets, they are not equally optional in broadband regimes with energetic high-wavenumber content. The Fourier symbol of $Q$ grows as $(\sqrt{|k_1|} - \sqrt{|k_2|})^2$ for scale-asymmetric arguments, so the $Q$-pieces contribute strongly to four-wave interactions involving a high-$k$ daughter mode. Numerical experiments at high spatial resolution exhibit a clean exponential growth of the high-$k$ spectral tail under the simplified model \eqref{main_eqn}, with no analogous instability under the full model \eqref{eq:full_reduction}; the $Q$-terms thus provide a high-wavenumber regularization absent from the simplified model. 

The Benjamin--Feir simulation of figure~\ref{fig:BF} illustrates this concretely. At $N = 1024$, \eqref{main_eqn} is stable at least to $t = 1250$ (tested maximum), well beyond several recurrence cycles; at $N = 2048$, with all other parameters unchanged, the spectral tail above $k \approx 200$ grows exponentially from the numerical floor while the carrier-band evolution remains in close agreement with the reference solvers. The integration fails at $t \approx 258$ (the time moment shown in figure \ref{fig:BF}b), just at the first recurrence peak, with $\max|\eta_x| \approx 0.29$ shortly before failure. The same exercise repeated with \eqref{eq:full_reduction} produces no instability at either resolution.

In high-resolution simulations of steep broadband regimes, the full model  \eqref{eq:full_reduction} is therefore the safer choice. For most analytical work, the Q-terms are essentially invisible and may be safely neglected.

\subsection{Breather solutions}\label{sec:breathers}
The sign-support property of the kernel (appendix~\ref{app:kernel}) has a dynamical consequence: the two-parameter ansatz $\hat u(k,t)=\hat B(k)\,\mathrm{e}^{-\ii(kV+\Omega)t}$ for $k>0$, subject to the reality condition $\hat u(-k,t)=\overline{\hat u(k,t)}$, closes exactly under \eqref{eq:full_reduction}. The equation therefore admits exact breather solutions. The stationary profile $\hat B(k)$ solves the nonlinear eigenproblem $(Vk+\Omega-\sqrt{k})\,\hat B=\ii\,\hat{R}[\hat B]$ for $k>0$, with $R$ the cubic right-hand side of \eqref{eq:full_reduction}, and is
computed by the Petviashvili method in the same way as for the super compact equation \citep{DyachenkoKachulinZakharov2017}. On the domain $L=2\pi$ with $N=2048$ points, the iteration converges to relative residuals of order $10^{-13}$ across the tested steepness range ($\max|\eta_x|$ up to $0.24$). The constructed
profile propagates without envelope deformation under \eqref{eq:full_reduction}, with relative $L^{2}$ shape error below
$10^{-9}$ over $100$ carrier periods at $\max|\eta_x|=0.24$ ($V=0.0625$, $\Omega=4.08$, giving $k_0=64$, $\max|\eta|\approx 0.0035$). In tested runs, the breather shape $\eta$ differs from the super compact breather at matched $(V,\Omega)$ by less than $0.5\%$ in relative $L^{2}$ norm (both reconstructed up to $\mathcal{O}(\epsilon)$). The simplified model \eqref{main_eqn} loses the exact sign support with the neglected $Q$-terms and supports breathers only approximately, with its envelope shape changing slightly over long times.

\section{Conclusions}\label{sec:conclusions}

We have derived a real-variable unidirectional reduction of the 
deep-water gravity-wave problem at $\mathcal{O}(\epsilon^2)$. The reduction yields a broadband evolution equation \eqref{eq:full_reduction} and its compact approximation \eqref{main_eqn}. Both retain full linear dispersion, admit the third-order Stokes wave as an exact solution, and reduce to Dysthe's envelope equation under a multiple-scales expansion. The nonlocal mean-flow term emerges automatically from the operator structure. Additionally, \eqref{eq:full_reduction} admits exact breather solutions, while \eqref{main_eqn} admits them only approximately. Both models agree closely with the super compact equation of \citet{DyachenkoKachulinZakharov2017} across narrow-band and broadband configurations.

The $Q$-terms of \S\ref{sec:simplified_model} carry the sub-leading $\mathcal{O}(\delta^2)$ corrections that the compact form \eqref{main_eqn} discards. Although asymptotically small on narrow-band content, they are structurally important at broad bandwidth, and provide the high-wavenumber regularization that suppresses the short-wave instability of \eqref{main_eqn}. For analytical work and moderate-amplitude simulations, the compact form is preferable; the full equation \eqref{eq:full_reduction} is the more robust choice at high spatial resolution and in high-steepness, broadband regimes.

The super compact equation is Hamiltonian, derived via Krasitskii-style canonical transformations in complex Fourier 
variables; \eqref{eq:full_reduction} and \eqref{main_eqn}, derived here using real variables, are not. Whether a Hamiltonian unidirectional reduction admits a real-variable operator form remains an open question. Such a construction would require either a further canonical transformation involving non-elementary operators or, in the variable $u$, an equally complex Poisson structure, more intricate than $J = -\tfrac{1}{2}A$ of the linear reduction.

The closed-form real-variable structure of \eqref{main_eqn} makes it well-suited to analytical study of focusing wavepackets, rogue-wave precursors, and the statistics of one-dimensional broadband random wave trains, where the absence of an envelope ansatz removes a restriction of Dysthe-class models.

\begin{bmhead}[Acknowledgements.]
The author expresses sincere thanks to Pearu Peterson for implementing the original conformal-mapping Euler solver and the periodic finite-difference (FD) scheme for spatial derivatives, and to Martin Laasmaa for modernizing the FD code execution for contemporary Python environments. Helpful discussions with Jaan Kalda, Dmitri Kartofelev, and Andrus Salupere are also gratefully acknowledged.
\end{bmhead}

\begin{bmhead}[Declaration of AI Use.]
During the preparation of this manuscript in April--May 2026, the author utilized generative AI tools (Claude Opus 4.7 and Gemini 3 Flash) for language editing and proofreading in some parts of the manuscript. All AI-assisted text was reviewed and revised by the author, who takes full responsibility for the final content. Claude was additionally used for exploratory symbolic computation, with specific use in appendix~\ref{app:kernel} documented therein. All other mathematical derivations, numerical results, and scientific conclusions remain entirely the author's own work.
\end{bmhead}

\begin{bmhead}[Funding.]
This research received no specific grant from any funding agency, commercial or not-for-profit sectors.
\end{bmhead}

\begin{bmhead}[Declaration of interests.]
The author reports no conflicts of interest.
\end{bmhead}

\begin{bmhead}[Data availability statement.]The simulation code and numerical data that support the findings of this 
study are available from the author upon request.\end{bmhead}

\begin{appen}
\renewcommand{\theHsection}{\Alph{section}}
\section{Fourier kernel of $\mathcal{H}_4$}\label{appA}
In Fourier space, for $\hat{f}(k)=(2\pi)^{-\frac{1}{2}}\int f(x)\,\mathrm{e}^{-\ii k x}\,\mathrm{d}x$, the quartic part of the Hamiltonian \eqref{hamiltonian_uv} is written as
\begin{equation}\label{H4_fourier}
  \mathcal{H}_4 = -\frac{\epsilon^2}{2\pi} \!\int\! K(k_1,k_2,k_3,k_4)\,
  \hat{u}(k_1)\hat{u}(k_2)\hat{v}(k_3)\hat{v}(k_4)\,\delta\!\Big(\textstyle\sum_i k_i\Big)
  \prod_i \mathrm{d}k_i,
\end{equation}
where $\hat{u},\hat{v}$ are the Fourier transforms of $u,v$. A direct evaluation, 
symmetrized in $k_1\leftrightarrow k_2$ and $k_3\leftrightarrow k_4$, yields 
(with $\sigma_i \equiv \mathrm{sgn}(k_i)$):
\begin{equation}\label{eq:K-closed}
  K(k_1,k_2,k_3,k_4) = \tfrac{1}{8}|k_3||k_4|\,(1+\sigma_1\sigma_2)
  \!\left[\,2(|k_3|+|k_4|) \;-\!\!\sum_{\substack{i\in\{1,2\}\\ j\in\{3,4\}}}\!\!|k_i+k_j|\,\right].
\end{equation}
The prefactor $(1+\sigma_1\sigma_2)$ eliminates every configuration in which $k_1$ and $k_2$ have opposite signs. We define the $m$-$n$ wavenumber-sign sector as $m$ of the four wavenumbers carrying one sign and $n$ the other. The 4-0 sector is empty on the momentum shell. On the remaining configurations, momentum conservation
$\sum_i k_i = 0$ forces the bracket in \eqref{eq:K-closed} to evaluate to
\begin{equation}\label{bracket_sectors}
  2(|k_3|+|k_4|) - \!\!\sum_{i,j}\!|k_i+k_j| \;=\;
  \begin{cases}
    0 & \sigma_1=\sigma_2, \quad\sigma_3 = -\sigma_4\\
    4\min(|k_1|,|k_2|,|k_3|,|k_4|) & \sigma_1=\sigma_2, \quad\sigma_3 = \sigma_4
  \end{cases}
\end{equation}
so that on the momentum shell
\begin{equation}\label{eq:K-sectorred}
  K(k_1,k_2,k_3,k_4)\Big|_{\sum_i k_i = 0} \;=\;
  \begin{cases}
    |k_3||k_4|\,\min(|k_1|,|k_2|,|k_3|,|k_4|) & \sigma_1=\sigma_2=-\sigma_3=-\sigma_4 \\[2pt]
    0 & \text{otherwise.}
  \end{cases}
\end{equation}
The kernel is non-zero only in the aligned part of the 2-2 wavenumber-sign sector, in which the two $u$
wavenumbers ($k_1$, $k_2$) share one sign and the two $v$ wavenumbers ($k_3$, $k_4$) share the other. On the resonant manifold $K=|k_1k_2|\,\min(|k_1|,|k_2|)$ and, expressed in normal variables, it agrees with the standard deep-water 4-wave kernel of~\citet{Krasitskii1994}; off resonance it realizes a gauge-equivalent form  \citep[see also][]{DyachenkoKachulinZakharov2017}.


\section{Fourier kernel of the cubic source $r$}
\label{app:kernel}

For real $u(x)$, with Fourier transform $\hat u(k)=(2\pi)^{-\frac{1}{2}}\int u(x)\,\mathrm{e}^{-\ii k x}\,\mathrm{d}x$, we define the kernel $M(k_1,k_2,k_3)$ of \eqref{q+pt_2} by
\begin{equation}\label{eq:kernel_def}
\hat r(k) \;=\; \frac{1}{2\pi}\int M(k_1,k_2,k_3)\,\hat u(k_1)\hat u(k_2)\hat u(k_3)\,
\delta(k-k_1-k_2-k_3)\,\mathrm{d}k_1\,\mathrm{d}k_2\,\mathrm{d}k_3.
\end{equation}
The operator symbols are
$\hat{A}=\ii\sgn(k)\sqrt{|k|}$, $\hat{H}=-\ii\sgn(k)$,
$\hat{\partial_x}=\ii k$, and $\hat{AH}=\sqrt{|k|}$.
Writing $r = r_{1a}+r_{1b}+r_{2a}+r_{2b}+r_3+r_4+r_5$ for the
seven terms of~\eqref{q+pt_2} and assigning $k_3$ to the
outermost factor of $u$ (the $u_x$ in $r_1$, the $Hu$ in $r_2$,
the bare $AHu$ in $r_4$) with $k_1,k_2$ in the inner pair, the
direct evaluation of~\eqref{eq:kernel_def} yields
\begin{subequations}\label{eq:per_term}
\begin{align}
M_{1a} &= +\sgn(k_1)\sgn(k_2)\sgn(k_1{+}k_2)\,k_3\,\sqrt{|k_1 k_2|}, \\
M_{1b} &= -\sgn(k_2)\sgn(k_1{+}k_2)\,k_1 k_3, \\
M_{2a} &= -\sgn(k_1)\sgn(k_2)\sgn(k_3)\sgn(k_1{+}k_2)\,|k|\sqrt{|k_1 k_2|}, \\
M_{2b} &= +\sgn(k_2)\sgn(k_3)\sgn(k_1{+}k_2)\,|k|\,k_1, \\
M_3 &= +2\,\sgn(k_1)\sgn(k_2)\,|k|\,\sqrt{|k_1 k_3|}, \\
M_4 &= -2\,\sgn(k_1)\sgn(k_2)\,|k_1{+}k_2|\,\sqrt{|k_1 k_3|}, \\
M_5 &= -\tfrac{1}{2}\,\sgn(k_1)\sgn(k_2)\sgn(k_3)\,(k_1{+}k_2)|k_1{+}k_2|,
\end{align}
\end{subequations}
where throughout this appendix $k\equiv k_1+k_2+k_3$. The full kernel is
$M = M_{1a}+M_{1b}+M_{2a}+M_{2b}+M_3+M_4+M_5$, and the symmetric kernel
$M^{\mathrm{sym}}$ is its average over the six permutations of
$(k_1,k_2,k_3)$. After simplification, the symmetric kernel of $r$ is
\begin{equation}\label{eq:Mclosed}
  M^{\mathrm{sym}}(k_1,k_2,k_3) \;=\;
  \begin{cases}
   F(k_1, k_2, k_3)
      & \text{for}\,\,\,\sgn(k_1)+\sgn(k_2)+\sgn(k_3) \;=\; \sgn(k), \\[4pt]
    0 & \text{otherwise,}
  \end{cases}
\end{equation}
where
\begin{equation}\label{F_closed}
     F(k_1, k_2, k_3)=-\tfrac{2}{3}\,\min(|k_1|,|k_2|,|k_3|,|k|)\,
    \Bigl[\omega_1\omega_2 + \omega_2\omega_3 + \omega_3\omega_1
          - \tfrac{1}{2}\bigl(|k_1|+|k_2|+|k_3|-|k|\bigr)\Bigr],
\end{equation}
and $\omega_i=\sqrt{|k_i|}$.

The closed-form expression \eqref{eq:Mclosed} was established with the assistance of Claude Opus 4.7 \citep{Claude} and
independently verified in Maple \citep{Maple2019}.

\section{Numerical methods}\label{sec:app_numerics}
All solvers except CM operate on the same uniform grid
$x_j = jL/N$, $j = 0, \ldots, N-1$, with periodic boundary conditions on $[0, L)$. The Hilbert transform $H$ and the pseudo-differential operator $A$ are evaluated pseudo-spectrally via the FFT. Spatial derivatives are calculated using a periodic finite difference scheme, except for SC. Each integrator advances its own internal time step subject to the prescribed tolerances and is queried at the fixed output interval $\Delta t$.

The super compact equation is integrated pseudo-spectrally with a $2/3$ dealiasing rule, using an adaptive Dormand--Prince scheme (\texttt{dopri5}, $r_{\mathrm{tol}} = 10^{-10}$, $a_{\mathrm{tol}} = 10^{-12}$) in the interaction picture $\hat C(k,t) = \exp[+\mathrm{i}\omega(k)\,t]\,\hat c(k,t)$, which removes the linear oscillation analytically. The physical surface elevation $\eta(x,t)$ is recovered through the second-order canonical transformation \citep{DyachenkoKachulinZakharov2017}. The third-order $\eta$--$\phi$ system is integrated similarly with \texttt{dopri5} on the joint state vector $(\eta, \phi)^\top$. The conformal-mapping solver \citep{DyachenkoEtAl1996} operates on a conformal grid $\xi_j=j2\pi/N$ with $\bigl(x(\xi,t),\,\eta(\xi,t)\bigr)$ as the parametrized free surface, and is advanced using the \texttt{vode} multistep integrator ($r_{\mathrm{tol}} = 10^{-14}$, $a_{\mathrm{tol}} = 10^{-13}$) on the conformal state vector $(\eta(\xi,t), \phi(\xi,t))^\top$.
The unidirectional model \eqref{main_eqn} is integrated as a single evolution equation for the canonical variable $u$ with \texttt{dopri5} ($r_{\mathrm{tol}} = 10^{-10}$, $a_{\mathrm{tol}} = 10^{-12}$). The physical surface elevation is recovered at output time through the inverse transformation \eqref{eq:can_eta}. 

Initial conditions are imposed in the
natural state variables of each formulation: physical $(\eta, \phi)$ for the third-order solver, the canonical variable $u_0$ for the unidirectional model (using the map \eqref{transform_u(eta)}), conformal $(\eta(\xi,0), \phi(\xi,0))$ for CM, and the spectral amplitude $\hat c(k)$ for the super compact equation (using the inverted second-order transformation). For CM, the initial conformal parameterization $x(\xi,0)$ is obtained by
fixed-point iteration on the prescribed profile $\eta_0$. The initial condition is given only as $\eta_0(x)$, and for the bidirectional models, the reduction \eqref{reduction} with \eqref{transform_u(eta)} is used for calculating the initial velocity potential $\phi_0(x)$.

The solver state is written to an HDF5 file with one frame per output interval, together with diagnostic time series for conservation laws. All simulations were performed in double precision.

\end{appen}
\bibliographystyle{jfm}
\bibliography{paivos_bib}

\end{document}